# Controlled Fabrication of Metallic Electrodes with Atomic Separation

A. F. Morpurgo and C. M. Marcus
*Department of Physics, Stanford University, Stanford, California 94305-4060*

D.B. Robinson
*Department of Chemistry, Stanford University, Stanford, California 94305-5080*

**Abstract**

We report a technique for fabricating metallic electrodes on insulating substrates with separations on the 1 nm scale. The fabrication technique, which combines lithographic and electrochemical methods, provides atomic resolution without requiring sophisticated instrumentation. The process is simple, controllable, reversible, and robust, allowing rapid fabrication of electrode pairs with high yield. We expect the method to prove useful in interfacing molecular-scale structures to macroscopic probes and electronic devices.

Rapid advances in the ability to manipulate (1-3) and measure (5-7) matter at the level of single atoms and molecules suggest that future technology may allow the fabrication of electronic devices whose core consists of one or a few molecules. This possibility offers important technological advantages beyond a simple reduction in size, as single molecules can be designed and synthesized to perform a variety of specific electronic functions including molecular switches (8), rectifiers (9), magnetic and optically bistable systems (10), and even molecular transistors (11), allowing electronic functionality to be incorporated into chemical synthesis. However, what currently limits the systematic investigation of nanometer-scale electronic elements as well as their use as a viable technology (i.e. molecular electronics (12)) is the absence of a simple means of interfacing very small objects such as single molecules to macroscopic structures and devices.

At present, experiments probing the electrical properties of single atoms or molecules require either sophisticated techniques based on scanning probe microscopy, or special contacting schemes which often limit experimental flexibility. The latter is illustrated by the clever recent experiments measuring the electrical conductance of benzene-dithiol molecules using mechanical break junctions to provide two metallic contacts (13). This approach works well but is not readily adapted to include electrostatic gates, a feature that would broaden the experimental possibilities. On the other hand, even the best conventional lithographic methods (14) can not controllably produce electrodes separated by a few nanometers or less, which are necessary to contact most molecules of interest.

In this paper, we report a technique that readily allows the fabrication of pairs of metallic electrodes with atomic scale separation on an insulating substrate. The crucial innovation of this technique, which is based on standard lithography combined with electrochemical deposition, is active monitoring and control of the separation between electrodes *during the* *fabrication process*. The simplicity and robustness of the technique suggests that large-scale implementation for the purpose of nanoelectronic device fabrication should be possible.

The technique involves two main steps, as illustrated in Fig. 1. First, metallic electrodes are prepared using conventional microfabrication (Fig. 1(A)). The separation between electrodes at this stage is not critical. In the second step, metal is electrodeposited on top of the existing pattern from an electrolyte solution (Fig. 1(B)). This results in an increase in the size of the electrodes, and hence a decrease in their separation (Fig. 1(C)). By measuring the electrical resistance between the two electrodes, we are able to monitor their separation once this distance becomes very small. In practice, monitoring the resistance signal allows controlled deposition with atomic-scale resolution. The process can be reversed to controllably widen gaps with similar accuracy. In fact, one can deposit until the electrodes are in contact and subsequently electrodissolve the metal to reopen the gap.

Examples of electrode pairs fabricated by this technique are shown in Fig. 2. Coarsely spaced Ti/Au (15 nm/35 nm) electrodes were patterned on a thermally oxidized silicon substate electron-beam lithography and lift-off (15). Initial spacings were in the range 50 – 400 nm. Samples were then placed in an aqueous solution consisting of 0.01 M potassium cyanaurate ($KAu(CN)_2$), and a buffer (pH 10) composed of 1 M potassium bicarbonate ($KHCO_3$) and 0.2 M potassium hydroxide. In the deposition reaction, the cyanaurate ion accepts an electron from the electrode and liberates the cyanide ligands, leaving a neutral gold atom at the surface. A gold pellet, 2 – 3 mm in diameter, was immersed in the solution to act as a counterelectrode. Thin gold wires (25 μm diameter, with ~ 3 – 4 mm of length in contact with the solution) were used to connect the patterned electrodes and the counterelectrode to the electrical circuit shown in Fig. 1(B). The complete circuit simultaneously serves to drive the electrodeposition process as well as monitor the interelectrode resistance.

During electrodeposition, a voltage bias of -0.5 to -0.6 V was applied to both electrodes relative to the counterelectrode, inducing a deposition current of 2 to 3 µA, resulting in gold plating at a lateral rate of ~ 1 Å/s. A number of values for the deposition current were used successfully and no effort has been made yet to optimize the process. The resistance between the two electrodes was measured by applying a 4 mV ac bias at 1 Hz across the electrodes and measuring the ac "monitor" current through a 1kΩ series resistor using a lock-in amplifier (Fig. 1(B)) (16).

Three phases of electrodeposition corresponding to different ranges of electrode separation can be identified from the time evolution of the monitor current. In the first phase, when the electrodes are far apart, the ac monitor current (~ 20 nA) is small and roughly constant (Fig. 3(A)). This current is proportional to the immersed surface area of the electrodes (dominated by the surfaces of the 25 µm gold wires) and results from the ac modulation of the dc deposition current. The second phase is marked by the sudden increase of the monitor current (Fig. 3(A), inset). At this point the electrodes are already very close, less than 5 nm, as shown below. The additional current observed in this phase is presumably due to direct tunneling between the contacts, enhanced by the screening effect of ions in the gap, which reduces the height of the tunnel barrier (17). The third phase, when the contacts finally touch, is marked by a sudden jump in the monitor current, followed by its saturation at a value given by the applied voltage divided by the ~ 1kΩ series resistance.

During the second phase of electrodeposition, when the electrodes are very close together but not yet touching, the monitor current is extremely sensitive to electrode distance, enabling control of the separation on an atomic scale. This is illustrated by Fig. 3(C), in which the deposition rate was reduced by a factor of 50 (by reducing the deposition current to ~ 50 nA) following the increase in monitor current. Using such small deposition currents allows the first atom(s) connecting the two electrodes to be resolved. These first atoms bridging the gap between the electrodes give rise to jumps in the monitor current corresponding to steps of ~ $2e^2/h$ in the conductance (Fig. 3(C), left inset), as expected for a single gold atom (7), which has a single electronic valence state available for conduction. Typically, only one or two steps of this magnitude are observed, followed by larger jumps presumably originating from clusters of atoms close to the contact point re-assembling themselves into more energetically favorable configurations. These steps are similar to those seen in electrodeposited Cu nanowires made using an STM (18).

The appearance of sharp steps in the monitor current associated with atomic conduction allows two important conclusions to be drawn. First, that this controlled deposition technique has atomic-scale resolution, so that it can be used to fabricate electrodes with ~1 nm separation reliably. Second, the steps unambiguously mark when the two electrodes touch; if electrodeposition is stopped at any earlier stage it is assured that the electrodes are not in direct contact.

We have fabricated many pairs of electrodes, stopping electrodeposition when the increase in the monitor current was first detected, and subsequently imaged the samples using a scanning electron microscope (SEM). Neither the SEM (Fig. 2) nor atomic force microscopy could resolve gap clearly, but placed consistent upper limits of 5 nm on the separation. Electrical resistances between such pairs of electrodes (measured using a 0.1 V bias in air after the fabrication) were between 1 and 30 GΩ, and in a few cases as low as 0.5 GΩ, whereas unplated electrodes on the same substrate had resistances above several hundred gigaohms, limited by the noise of the measurement. These values are consistent with electronic tunneling through a gap of roughly 1 nm (19).

We emphasize that no tuning of fabrication parameters was needed to achieve the present results, demonstrating the robustness of the technique. Alternative strategies have been reported recently (20) capable of feature sizes approaching those reported here, however, the present method offers several advantages including extremely small gaps, high yield (approaching 100%) at gap sizes down to ~1nm, relatively short fabrication time, and simple, readily available instrumentation.

Because this process can employ techniques and instruments that are currently in use in a variety of industries, including microelectronics manufacturers (deep-uv lithography and electroplating), it may be readily realized in an industrial setting. Note also that electronic feedback can easily be incorporated into the monitoring scheme, allowing the electrodeposition rate to be adjusted as a function of the resistance between electrodes and then stopped at a specified separation. This type of feedback control lends itself to parallel operation and provides a means of fabricating many structures at the same time.

We thank C.E.D. Chidsey for useful discussions and for the use of equipment in his laboratory. Research supported by the National Science Foundation PECASE program, DMR-9629180-1, and the Stanford Center for Materials Research, an NSF-MRSEC.

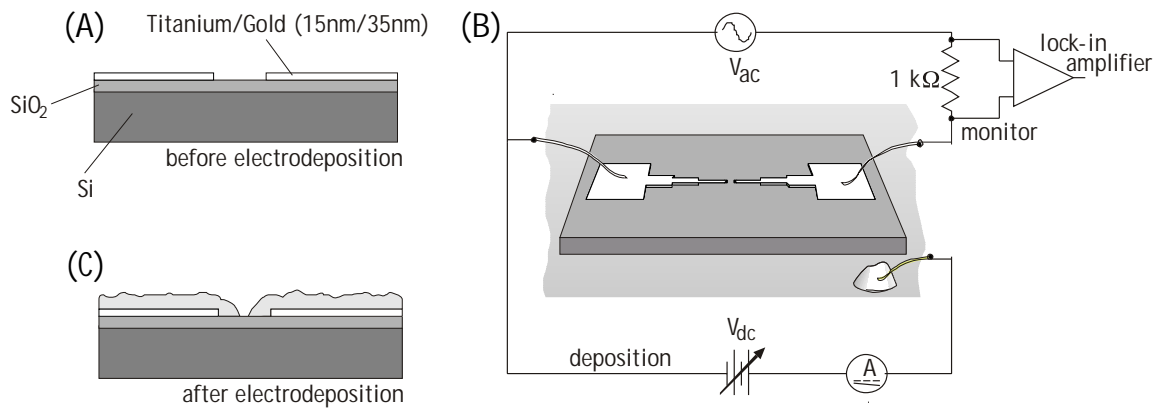

**Figure 1**
The fabrication of nanoelectrodes consists of two main steps: (A) Electrodes with large separation are fabricated by conventional lithography. (B) Metal is electrodeposited onto the electrodes, reducing their separation. $V_{dc}$ controls electrodeposition while $V_{ac}$ is used to monitor the conductance and thus the separation between the electrodes. Reversing $V_{dc}$ allows material to be removed rather than deposited. (C) When deposition is stopped before the electrodes touch, separations on the 1 nm scale are obtained reproducibly.

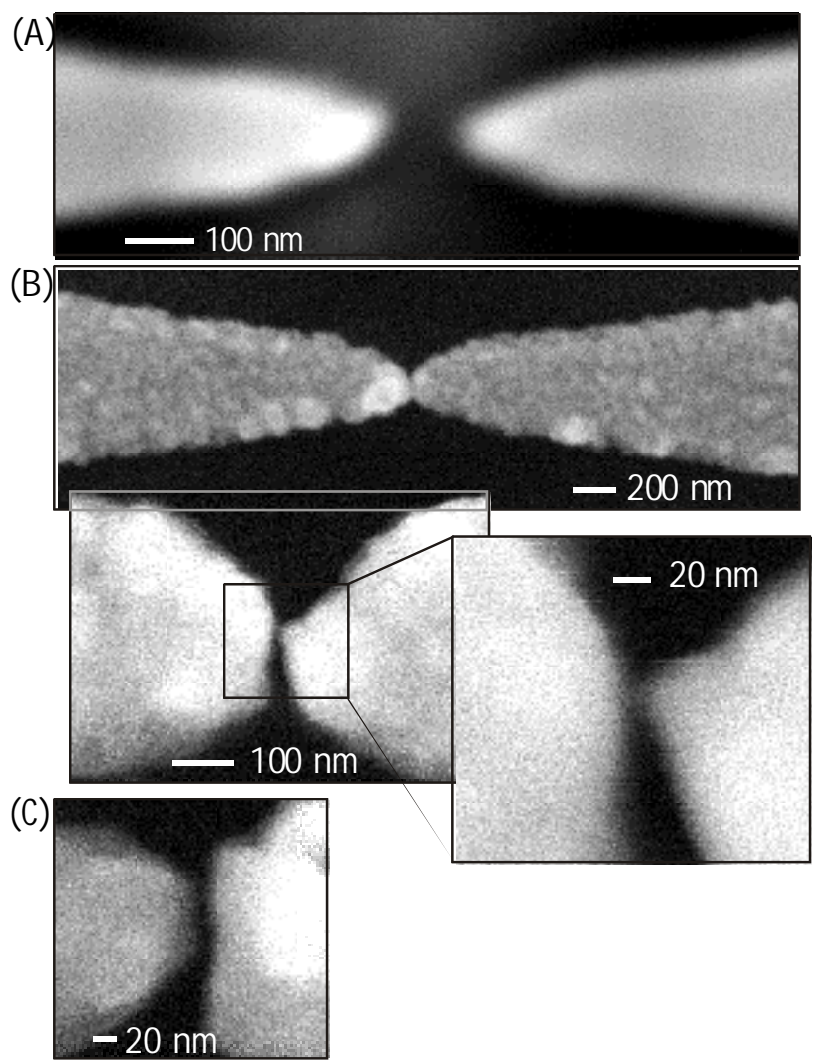

**Figure 2**
Scanning electron microscope (SEM) images before and after electrodepostion (scale bars show dimensions). (A) Electrodes before electrodeposition. (B) Electrodes after electrodeposition. The resolution of the SEM is 5 nm, not sufficient to resolve the gap. (C) Electrodes in which the gap was re-opened by electrodissolution, by reversing $V_{dc}$ following an intentional short-circuiting (contacting) in a previous electrodeposition process.

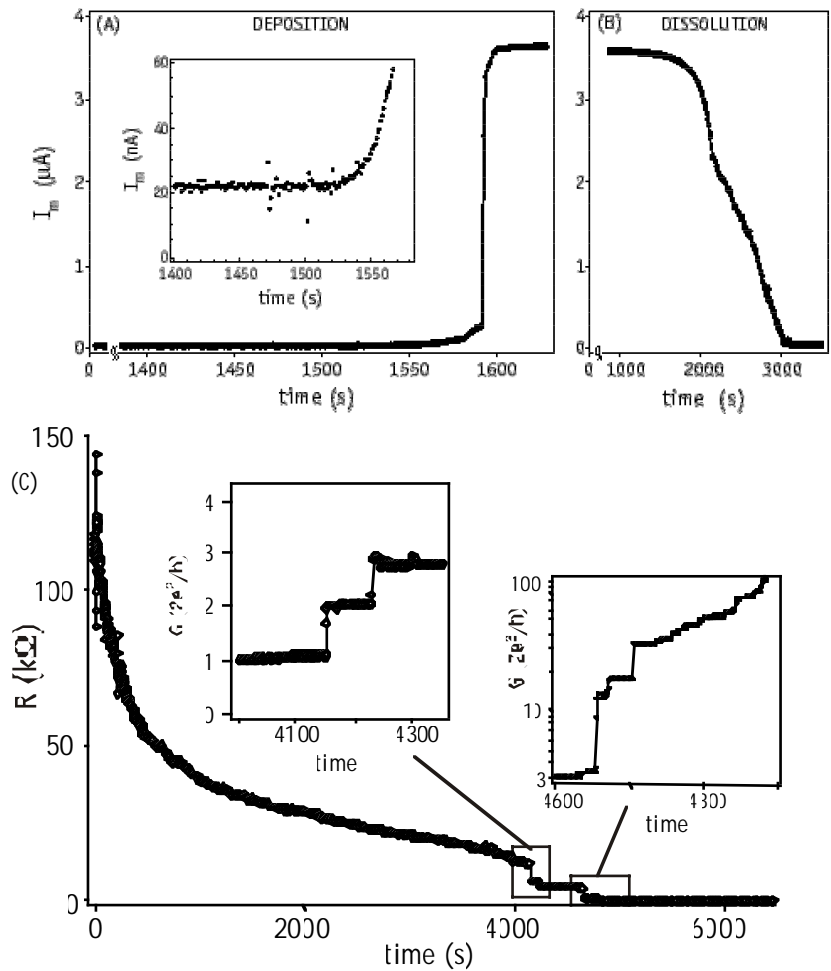

**Figure 3**
Time evolution of the ac monitor current during rapid electrodeposition (A) and electrodissolution (B). Three phases of electrodeposition can be identified (1) In this example, for times before ~ 1540 s, a small ac monitor current is measured when the electrodes are well separated, (2) For times between ~ 1540 s and 1590 s, a continuously increasing monitor current appears as the electrodes approach one another at the nm scale, (3) At ~ 1590 s, a sudden jump in the monitor current is observed as the electrodes make contact, followed by saturation. The time evolution is reversed for dissolution.

(C) Time evolution of the resistance R between electrodes for slow deposition (roughly 50 times slower than in Fig. 3(A)). Conductance steps close to $2e^2/h$ (the expected value for Au atoms) are visible in the left inset. Following initial contact, plateau-like features and steps in the conductance on the order of a few $e^2/h$ persist as the contact between electrodes continues to increase in size at the atomic scale (right inset).